\documentclass[%
 aip, pop,
 amsmath,amssymb,
 reprint,%
]{revtex4-1}

\usepackage{graphicx}
\usepackage{dcolumn}
\usepackage{bm}
\usepackage[mathlines]{lineno}

\usepackage[utf8]{inputenc}
\usepackage[T1]{fontenc}
\usepackage{mathptmx}
\usepackage{etoolbox}
\usepackage{orcidlink}   
\usepackage{booktabs}
\usepackage{comment}
\usepackage{ragged2e}

\makeatletter
\def\@email#1#2{%
 \endgroup
 \patchcmd{\titleblock@produce}
  {\frontmatter@RRAPformat}
  {\frontmatter@RRAPformat{\produce@RRAP{*#1\href{mailto:#2}{#2}}}\frontmatter@RRAPformat}
  {}{}
}%
\makeatother
\begin{document}

\preprint{AIP/123-QED}

\title{Non-thermal electron cyclotron emission during runaway plateau in tokamak disruptions from a highly anisotropic dielectric tensor}
\author{Yeongsun Lee\orcidlink{0000-0003-4474-416X}}
\affiliation{Seoul National University, Seoul, South Korea}
\affiliation{Nuclear Research Institute for Future Technology and Policy, Seoul National University, Seoul, South Korea}

\author{Kikyung Park\orcidlink{0009-0001-1397-2701}}
\affiliation{Seoul National University, Seoul, South Korea}

\author{Tchanou Park\orcidlink{0009-0008-1471-2013}}
\affiliation{Seoul National University, Seoul, South Korea}

\author{Gunsu Yun\orcidlink{0000-0002-1880-5865}}
\affiliation{Department of Physics, POSTECH, Pohang, Republic of Korea}
\affiliation{Division of Advanced Nuclear Engineering, POSTECH, Pohang, Republic of Korea}

\author{Yong-Su Na\orcidlink{0000-0001-7270-3846}}
\affiliation{Seoul National University, Seoul, South Korea}

\author{Jong-Kyu Park\orcidlink{0000-0003-2419-8667}$^*$}%
\email{jkpark@snu.ac.kr}
\affiliation{Seoul National University, Seoul, South Korea}

\date{\today}

\begin{abstract}
During the runaway plateau phase in a tokamak, a cold background electron temperature of $\mathcal{O}(1 \, \mathrm{eV})$ forbids the onset of kinetic instability due to strong collisional damping. Nevertheless, non-thermal ECE anomalies at the level of 100 eV to keV have been observed in this phase without externally injected waves. To explore this, we characterize a highly anisotropic hot plasma medium with a Gaussian pitch-angle distribution. We derive an analytic hot plasma dielectric tensor, yielding direct expressions for the non-thermal emission coefficients and the kinetic instability drive rate. These analytic forms are verified against the KIAT and SYNO codes at small pitch-angle spread and are numerically complemented at large pitch-angle spread. Using the method of images, we define a fictitious global temperature of the entire plasma medium as measured by a horizontal ECE system. Because this representative medium temperature can exceed the keV level, the radiative temperature measured under incomplete wall reflection can be highly non-thermal without invoking any kinetic instability. This interpretation provides a conceptual basis for quantitative validation against experimental ECE measurements under realistic conditions.
\end{abstract}

\maketitle

\section{Introduction}
The hot plasma dielectric tensor describes the response of a plasma medium arising from resonant interactions between charged particles and electromagnetic fields. This response governs wave propagation, polarization, as well as amplification and damping processes \cite{Stix1992}. Therefore, a proper evaluation of the hot plasma dielectric tensor is essential for analyzing wave characteristics in any plasma medium. For example, the hot plasma dielectric tensor plays central roles in accounting for strongly enhanced electron cyclotron emission (ECE) intensity observed in tokamak experiments. Such observations are commonly interpreted by considering quasi-linear diffusion of the electron distribution driven by runaway-induced kinetic instabilities \cite{Harvey1993PoF, Liu2018NF} or externally-injected waves \cite{Harvey1993PoF,Votta2026PPCF}, followed by analysis of the ECE emitted from the resulting strongly scattered distribution or from a phenomenological pitch-angle distribution \cite{Yu2026NF}.

However, there exists an important problem that must be explained without invoking such quasi-linear diffusion: the non-thermal ECE emission often observed in tokamak disruption experiments. In post thermal quench phase, the onset of kinetic instabilities without external wave injections is difficult due to the low background electron temperature \cite{Aleynikov2015NF}. For instance, non-thermal ECE signals of $0.1 - 1 \, \mathrm{keV}$ were measured during the runaway plateau with a $2 \, \mathrm{eV}$ background electron temperature in the DIII-D tokamak \cite{Hollmann2013NF}. A linear stability analysis showed no positive growths for that experiment in Ref. \cite{Aleynikov2015NF}, where the energy and pitch-angle distributions of energetic electrons were inferred from the experimentally measured data from Ref. \cite{Hollmann2013NF}. This suggests that energetic electrons may still dominate the ECE emission even with a highly anisotropic distribution such as a Gaussian pitch-angle distributions. One key clue is that L-mode plasmas considered in the previous works exhibited electron temperatures of several hundred eV \cite{Harvey1993PoF, Liu2018NF}, whereas the runaway plateau has a background temperature of only $\mathcal{O}(1 \, \mathrm{eV})$ \cite{Hollmann2013NF, Aleynikov2015NF}. This leads to a correspondingly large difference in optical thickness, which may play a key role in understanding the non-thermal feature in the runaway plateau. Motivated by this, in this work, we explore temperature anomalies under a disruption-relevant optically-transparent tokamak.

The central question is how one can define a temperature for a tokamak plasma medium containing an anisotropic runaway-electron population when a horizontal ECE system measures the corresponding radiation intensity. Once such a definition is established, the temperature anomalies observed during the disruption phase have a simple explanation: the representative temperature measured by the ECE system is merely higher than the thermal electron temperature. For example, an optically thin medium with the representative temperature of 1 keV radiates an ECE intensity corresponding to $100 \, \mathrm{eV}$, which requires no special interpretation. We present an analytic hot plasma dielectric tensor for a Gaussian pitch-angle distribution in Sec. II. In Sec. III, we apply the formulation to obtain analytic expressions for runaway-driven kinetic instability growth rates and non-thermal ECE coefficients. Using the method of images, we introduce a fictitious definition of global spectral non-thermal temperature of the tokamak plasma medium in the radiative balance. Although our model relies on several simplifying assumptions, this simplicity is sufficient for \textit{qualitatively} explaining how even runaway electron distributions that do not trigger kinetic instabilities can nevertheless produce anomalously enhanced ECE signals.

\section{Plasma dispersion relations}
In this section, we adopt cold and hot plasma dielectric tensors from Refs. \cite{Shafranov1967, bekefi}. A collisionless hot plasma dielectric tensor for Gaussian pitch-angle distributions characterizing runaway electron distributions is then obtained by applying the second exponential Weber integral. The integral procedure is analogous to that for Maxwellian distributions \cite{Stix1992} since the Maxwellian plasma distribution is also characterized by a Gaussian distribution.

\subsection{Hot electron distribution with a Gaussian pitch-angle distribution \label{ssec:fhot_theta0}}
We assume that hot electrons have a momentum spectrum $G(p)$ with the Gaussian angular spectrum \cite{Aleynikov2015NF}
\begin{equation}
    f_{hot}(p,\theta) = \frac{n_{hot}}{2\pi n_e} G(p) \frac{2\exp(-\frac{\theta^2}{\theta^2_0})}{\theta_0^2} \label{eq:f}
\end{equation}
where $f_{hot}$ is the hot electron distribution function, $p$ is normalized particle momentum, $\theta$ is pitch angle and $\int d^3p f_{hot} = n_{hot}/n_e$ is met. The remaining part is the Maxwellian, $f_M$ with $\int d^3p f_M = 1-n_{hot}/n_e \approx 1$. 

\subsection{Validity range of a Gaussian pitch-angle distribution \label{ssec:theta0_validity}}
Let $p_c$ be the critical momentum on which a friction force balances with an electric force in phase space and $\theta_0$ adjusted to ultra-relativistic electrons. Our conservative simplification of the Gaussian pitch-angle form with a constant $\theta_0$ underestimates perpendicular kinetic energy of mildly relativistic particles both in the non-runaway region $p \lesssim p_c$ and in the runaway region $p > p_c$, though to different degrees.

In the runaway region, the pitch-angle spread of the runaway distribution function (analogous to heat diffusion solution) \cite{Connor1975NF}, which was used for kinetic instability context in Ref. \cite{Fulop2006PoP}, depends on particle momentum such that $\theta_0^2(p) \propto p^{-1}$ when transport is slow and radiation is negligible. Accordingly, the underestimated spread in this region arises not from the anisotropic functional form (Gaussian), which remains valid, but from assuming too small $\theta_0$. This underestimation can be alleviated by scanning over a higher range of $\theta_0$ for the specific purpose of evaluating the non-thermal ECE if the absence of kinetic instability is guaranteed for a certain $\theta_0$. In the non-runaway region, the pitch-angle distribution could be far broader \cite{Rosenbluth1997NF, Hollmann2013NF}. If this region dominates the non-thermal ECE, varying $\theta_0$ alone would be insufficient, and a qualitatively different functional form such as an isotropic distribution may need to be introduced.

The Gaussian form would be applicable during the runaway plateau phase, but not necessarily throughout the entire disruption phase. For instance, when the runaway current decays collisionally, an electric field $E$ is close to the so-called critical electric field $E_c$ \cite{Rosenbluth1997NF, Aleynikov2015PRL} and the corresponding condition $p_c \equiv \frac{1}{\sqrt{E/E_c - 1}} \approx \mathcal{O}(1)$ implies the dominant contribution of non-runaway particles. However, in the controlled runaway plateau phase, a finite runaway avalanche growth ($E/E_c \gg 1$) is required for compensating their loss. If an electric field far exceeds the critical electric field and $p_c \ll 1$ is met, we expect that the pitch-angle distribution would be still anisotropic satisfying the condition
\begin{equation}
    \tau_{coll} \ll \tau_{ava} \approx \tau_{RE}
\end{equation}
where $\tau_{coll}$ is the collisional time scale required for reaching the runaway pitch-angle distribution to relax to its anisotropic Gaussian equilibrium, $\tau_{ava}$ is the runaway avalanche time (inverse growth rate) and $\tau_{RE}$ is the runaway confinement time. The former relation $\tau_{coll} \ll \tau_{ava}$ is evident since the avalanche is governed by "rare" knock-on collisions.

In DIII-D, interpreting the observed ECE spectrum under a single-particle assumption required a large pitch angle ($\theta \approx 0.8$) for electrons below $100\,\mathrm{keV}$, while high energy electrons exhibited a small pitch angle ($\theta \approx 0.2$) \cite{Hollmann2013NF}. This is consistent with the broadening of the $\theta$-distribution in the non-runaway region discussed above, and indicates that a Gaussian approximation alone may be inadequate when low-energy electrons dominate the non-thermal ECE.

\subsection{Cold and hot plasma dispersion relations}
The complete dielectric tensor $\overleftrightarrow{\varepsilon}$ consists of the Hermitian $\overleftrightarrow{\varepsilon}_H$ and anti-Hermitian $\overleftrightarrow{\varepsilon}_A$ parts,
\begin{equation}
    \overleftrightarrow{\varepsilon} = \overleftrightarrow{\varepsilon}_H + i \overleftrightarrow{\varepsilon}_A
\end{equation}
where $\overleftrightarrow{\varepsilon}_A=\overleftrightarrow{\varepsilon}_A^{coll}+\overleftrightarrow{\varepsilon}_A^{hot}$ consists of the collisional part $\overleftrightarrow{\varepsilon}_A^{coll}$ and the collisionless part $\overleftrightarrow{\varepsilon}_A^{hot}$.

Let $\omega$ be wave frequency, $\omega_c\equiv \frac{|e|B}{m}$ be electron cyclotron frequency, $\omega_{ci}\equiv \frac{|e|B}{M}$ be ion cyclotron frequency and $\omega_p\equiv \sqrt{\frac{ne^2}{m \varepsilon_0}}$ be plasma frequency. $\overleftrightarrow{\varepsilon}_H$ is written by the cold plasma dispersion relation \cite{Shafranov1967},
\begin{equation}
    \overleftrightarrow{\varepsilon}_H = \begin{bmatrix}
        \varepsilon_H & i g_H & 0 \\
        -i g_H & \varepsilon_H & 0 \\
        0 & 0 & \eta_H
    \end{bmatrix}, \label{eq:ten_eps_H}
\end{equation}
where $\varepsilon_H = 1 - \frac{\omega_p^2}{\omega^2 - \omega_{c}^2}$, $g_H = - \frac{\omega_c}{\omega} \frac{\omega^2_p}{\omega^2 - \omega^2_c}$, $\eta_H = 1 - \frac{\omega_p^2}{\omega^2}$. In Eq. \ref{eq:ten_eps_H}, we only consider electron contributions. This simplification is valid when $\omega^2 \gg \omega_c \omega_{ci}$ \cite{Akhiezer1975} and thereby suitable for capturing either $l=2$ extraordinary mode wave (X-mode) for ECE reconstruction or the whistler wave for kinetic instability analysis. For the magnetized plasma wave, the neglect of ions can raise an error as $\omega$ goes to the lower hybrid frequency. In a low frequency limit, however, a finite $\theta_0$ required for meaningful non-thermal ECE strongly regulates a kinetic drive \cite{Aleynikov2015NF}. Hence, such an error is negligible when we investigate a growth rate of the least stable waves under strong non-thermal ECE regime with a finite $\theta_0$.

An effect of collisions can be considered by replacing $\omega \to \omega+ i \nu_{ei}$ in the conductivity tensor\cite{Aleynikov2015NF}, i.e. $\overleftrightarrow{\varepsilon}_A^{coll} \approx \frac{1}{i \varepsilon_0 \omega} \Big(\frac{\partial}{\partial \omega} i\varepsilon_0 \omega (\overleftrightarrow{\varepsilon}_H - \overleftrightarrow{I}) \Big)|_{\omega = \omega} \cdot \nu_{ei}$, which leads to
\begin{equation}
    \overleftrightarrow{\varepsilon}_A^{coll} = \begin{bmatrix}
        \varepsilon_A^{coll} & i g_A^{coll} & 0 \\
        -i g_A^{coll} & \varepsilon_A^{coll} & 0 \\
        0 & 0 & \eta_A^{coll}
    \end{bmatrix},
\end{equation}
where $\nu_{ei}\equiv\frac{\sqrt{2} \ln \Lambda n_e Z_{eff} e^4}{12 \pi^{3/2} \varepsilon_0^2 m^{1/2} T_e^{3/2}}$ is collision frequency, $\varepsilon_A^{coll} = \frac{\nu_{ei}}{\omega} \frac{(\omega^2 +\omega_c^2)\omega_p^2}{(\omega^2 - \omega_c^2)^2}$, $g_A^{coll} =  2 \frac{\nu_{ei}}{\omega} \frac{\omega \omega_{c} \omega_p^2}{(\omega^2 - \omega_c^2)^2}$ and $\eta_A^{coll} = \frac{\nu_{ei}}{\omega} \frac{\omega_p^2}{\omega^2}$.

$\overleftrightarrow{\varepsilon}_A^{hot}$ is written by the collisionless hot plasma dispersion relation \cite{bekefi, Lee2025phD}, where 'collisionless' means a limit $\nu_{ei}\to0$ rather than $\nu_{ei}=0$,
\begin{equation}
    \overleftrightarrow{\epsilon}_A^{hot} = 2\pi^2 \frac{\omega_p^2}{\omega^2} \int  dp_\| dp_\perp \gamma \sum\limits_{l=-\infty}^\infty \overleftrightarrow{S}_l U \delta(\omega - k_\| v_\| - l\frac{\omega_c}{\gamma}), \label{eq:eps_hot}
\end{equation}
where
\begin{equation}
    \overleftrightarrow{S}_l = \begin{bmatrix}
        \beta_\perp^2 (\frac{l}{\lambda})^2 J_l^2 & i \beta_\perp^2 \frac{l}{\lambda}J_l J_l' & \beta_\| \beta_\perp \frac{l}{\lambda} J_l^2 \\
        -i \beta_\perp^2 \frac{l}{\lambda} J_l J_l' & \beta_\perp^2 J_l'^2 & -i \beta_\| \beta_\perp J_l J_l' \\
        \beta_\perp \beta_\| \frac{l}{\lambda} J_l^2 & i \beta_\perp \beta_\| J_l J_l' & \beta_\|^2 J_l^2.
    \end{bmatrix}.
\end{equation}
and
\begin{equation}
    U = \omega \sin\theta \frac{\partial f_0}{\partial p} + \frac{\omega\cos\theta - k_\| v}{p} \frac{\partial f_0}{\partial \theta} \label{eq:U}
\end{equation}
and $l$ is the mode number, the argument of Bessel function $J$ and its derivative $J'$ is $\lambda = \gamma \frac{k_\perp v_\perp}{\omega_c}$ and $f_0$ is the normalized electron distribution function, $f_0 = f_{hot} + f_M$ with $\int d^3p f_0=1$. $k$ is the wave vector, $\beta$ is normalized particle velocity, $\gamma$ is the relativistic factor and the subscripts $\|$ and $\perp$ represent parallel and perpendicular components to magnetic field, respectively.

\subsection{Analytic hot plasma dielectric tensor for Gaussian pitch-angle distributions}
Although $f_0 = f_{hot} + f_M$, we only consider $f_{hot}$ in forthcoming calculation for brevity.

Hot plasma dielectric tensor has an integral across the pitch-angle $\theta$. The Gaussian-$\theta$ distribution in $f_{hot}$ allows the leading order approximation of $\sin \theta \approx \theta$ and $\cos \theta \approx 1$, including the argument of $J_l$ and $J_l'$. An error from the truncation is negligible due to a rapid-decaying factor $\exp(-\theta^2/\theta_0^2)$ when the approximations are broken.

Substituting. \ref{eq:f} in Eq. \ref{eq:U} yields the following
\begin{equation}
    U \approx \Big( \omega \frac{d}{dp} \ln G - 2\frac{\omega - k_\| v}{p \theta_0^2} \Big) \theta f_{hot}. \label{eq:approx_U}
\end{equation}
Represent an argument of delta function as $p$ using
\begin{equation}
    \delta(\omega - k_\| v_\| -l \frac{\omega_c}{\gamma}) \approx - \frac{\gamma^3}{k_\| c -l\omega_{c} p} \delta (p-p_{res}). \label{eq:approx_del}
\end{equation}
where the resonance condition is given by
\begin{equation}
    \gamma_{res} = \frac{k_\|^2 c^2 + l^2 \omega_c^2}{l \omega \omega_c + \sqrt{l^2 \omega^2 \omega^2_c + (k_\|^2 c^2 -\omega^2) (k_\|^2 c^2 + l^2 \omega_c^2)}}.
\end{equation}
Note that $\theta$-dependence in $\delta(\omega - k_\| v_\| -l \frac{\omega_c}{\gamma})$ and $\gamma_{res}$ is the 2nd order, so negligible in the 1st order expansion.

The direct integration over $p$ in Eq. \ref{eq:eps_hot}, after plugging Eqs. \ref{eq:approx_U} and \ref{eq:approx_del} and variable transformation from $(p_\|, p_\perp)$ to $(p, \theta)$, leaves the $\theta$-integral such that
\begin{widetext}
\begin{equation}
    \overleftrightarrow{\epsilon}_A^{hot} = \pi \frac{\omega_p^2 n_{hot}}{\omega^2 n_{e}} \sum\limits_{l=-\infty}^\infty \frac{p_{res}\gamma^4_{res}}{l \omega_c p_{res} - k_\| c} \Big( \omega \frac{dG}{dp}(p_{res}) - 2\frac{\omega-k_\|c \beta_{res}}{p_{res} \theta_0^2}G(p_{res}) \Big) \int  d\theta \theta \overleftrightarrow{S}_l \frac{2}{\theta_0^2} \exp(- \frac{\theta^2}{\theta_0^2}).
\end{equation}

\begin{table}[h!]
    \centering
    \begin{tabular}{@{} c|c|c @{}}
        \toprule
        & The original integral & The second exponential Weber integral ($\Lambda \equiv \frac{a^2\theta_0^2}{2}$) \\
        \midrule
        $\mathcal{W}_{l,1}^{1,\cos \theta}$ & $\int_0^{\pi} \frac{2\exp(-\frac{\theta^2}{\theta_0^2})}{\theta_0^2} J_l^2(a\sin\theta) \sin\theta [1, \cos \theta] d\theta$ & $\int_0^{\infty} \frac{2\exp(-\frac{\theta^2}{\theta_0^2})}{\theta_0^2} J_l^2(a\theta) \theta d\theta = \exp(-\Lambda) I_l (\Lambda)$  \\
        $\mathcal{W}_{l,2}^{1,\cos \theta}$ & $\int_0^{\infty} \frac{2\exp(-\frac{\theta^2}{\theta_0^2})}{\theta_0^2} J_l(a\sin\theta)J'_l(a\sin\theta) \sin^2\theta [1, \cos \theta] d\theta$ &  $\int_0^{\infty} \frac{2\exp(-\frac{\theta^2}{\theta_0^2})}{\theta_0^2} J_l(a\theta)J'_l(a\theta) \theta^2 d\theta = \frac{a\theta_0^2}{2} \exp(-\Lambda) \Big(I'_l (\Lambda)-I_l (\Lambda)\Big)$  \\
        $\mathcal{W}_{l,3}^{1,\cos \theta}$ & $\int_0^{\pi} \frac{2\exp(-\frac{\theta^2}{\theta_0^2})}{\theta_0^2} J'^2_l(a\sin\theta) \sin^3\theta [1, \cos \theta] d\theta$ &  $\int_0^{\infty} \frac{2\exp(-\frac{\theta^2}{\theta_0^2})}{\theta_0^2} J'^2_l(a\theta) \theta^3 d\theta = \frac{\theta_0^2}{2} \exp(-\Lambda) \Big[ \Big(2\Lambda + \frac{l^2}{\Lambda} \Big)I_l (\Lambda) - 2 \Lambda I'_l (\Lambda)\Big]$  \\
        \bottomrule
    \end{tabular}
    \caption{\justifying
    The second exponential Weber integral and its variants. Superscripts 1 and $\cos\theta$ in $\mathcal{W}_{l,\nu}^{1,\cos \theta}$ with $\nu=1, 2, 3$ denote the corresponding factors $[1, \cos\theta]$ in the integrand of the original integral.
    }
    \label{tab:Weber}
\end{table}
We evaluate each tensor elements of the integral using the second exponential Weber integral \cite{Watson1922} (see Table \ref{tab:Weber})
\begin{equation}
    \int d\theta \Big[\Big(\frac{l \omega_c}{\gamma_{res} k_\perp c} \Big)^2 \frac{2\theta }{\theta_0^2} \exp(- \frac{\theta^2}{\theta_0^2}) J_l^2 (\lambda_{res})\Big] = \Big(\frac{l \omega_c}{\gamma_{res} k_\perp c} \Big)^2 \exp(-\Lambda) I_l (\Lambda), \label{eq:Weber1}
\end{equation}
\begin{equation}
    \int d\theta \Big[i \frac{l \omega_c}{\gamma_{res} k_\perp c} \beta_{res} \frac{2\theta^2}{\theta_0^2} \exp(- \frac{\theta^2}{\theta_0^2}) J_l (\lambda_{res}) J_l' (\lambda_{res})\Big] \approx \frac{i l \beta_{res}^2 \theta_0^2}{2} \exp(-\Lambda) \Big( -I_l (\Lambda) + I_l' (\Lambda) \Big), \label{eq:Weber2}
\end{equation}
\begin{equation}
    \int d\theta \Big[\frac{l \omega_c}{\gamma_{res} k_\perp c} \beta_{res} \frac{2 \theta}{\theta_0^2} \exp(- \frac{\theta^2}{\theta_0^2}) J_l^2 (\lambda_{res})\Big] \approx \frac{l \omega_c \beta_{res}}{\gamma_{res} k_\perp c}\exp(-\Lambda) I_l (\Lambda), \label{eq:Weber3}
\end{equation}
\begin{equation}
    \int d\theta \Big[\beta_{res}^2 \frac{2 \theta^3}{\theta_0^2} \exp(- \frac{\theta^2}{\theta_0^2}) J_l'^2 (\lambda_{res})\Big] \approx \frac{\beta_{res}^2 \theta_0^2}{2} \exp(-\Lambda) \Big( (2\Lambda + \frac{l^2}{\Lambda}) I_l (\Lambda) - 2 \Lambda I_l' (\Lambda) \Big), \label{eq:Weber4}
\end{equation}
\begin{equation}
    \int d\theta \Big[ i \beta_{res}^2 \frac{2\theta^2}{\theta_0^2} \exp(- \frac{\theta^2}{\theta_0^2}) J_l (\lambda_{res}) J_l'(\lambda_{res})\Big] \approx \frac{i \beta_{res}^2 \theta_0^2}{2} \frac{k_\perp c p_{res}}{\omega_c}\exp(-\Lambda) \Big(-I_l (\Lambda) + I_l' (\Lambda) \Big), \label{eq:Weber5}
\end{equation}
\begin{equation}
    \int d\theta \Big[\beta_{res}^2 \frac{2\theta}{\theta_0^2} \exp(- \frac{\theta^2}{\theta_0^2}) J_l^2 (\lambda_{res})\Big] = \beta_{res}^2 \exp(-\Lambda) I_l (\Lambda) \label{eq:Weber6}
\end{equation}
where $\Lambda = \frac{k_\perp^2 c^2 p_{res}^2 \theta_0^2}{2 \omega_c^2}$. The final dielectric tensor becomes
\begin{equation}
\begin{split}
    \overleftrightarrow{\epsilon}_A^{hot} = \pi \frac{\omega_p^2 n_{hot}}{\omega^2 n_{e}} &\sum\limits_{l=-\infty}^\infty \frac{p_{res}\gamma^4_{res}}{l \omega_c p_{res}- k_\| c} \Big( \omega \frac{dG}{dp}(p_{res}) - 2\frac{\omega-k_\|c \beta_{res}}{p_{res} \theta_0^2}G(p_{res}) \Big) \\
    &\times\exp(-\Lambda)\begin{bmatrix}
        \Big(\frac{l \omega_c}{\gamma_{res} k_\perp c} \Big)^2 I_l (\Lambda) &
        \frac{i l \beta_{res}^2 \theta_0^2}{2} \Big(I_l' (\Lambda) -I_l (\Lambda) \Big) &
        \frac{l \omega_c \beta_{res}}{\gamma_{res} k_\perp c} I_l (\Lambda) \\
        -\frac{i l \beta_{res}^2 \theta_0^2}{2} \Big(I_l' (\Lambda) -I_l (\Lambda) \Big) &
        \frac{\beta_{res}^2 \theta_0^2}{2} \Big( (2\Lambda + \frac{l^2}{\Lambda}) I_l (\Lambda) - 2 \Lambda I_l' (\Lambda) \Big) &
        \frac{i \beta_{res}^2 \theta_0^2}{2} \frac{k_\perp c p_{res}}{\omega_c} \Big(I_l (\Lambda) - I_l' (\Lambda) \Big) \\
        \frac{l \omega_c \beta_{res}}{\gamma_{res} k_\perp c} I_l (\Lambda) &
        - \frac{i \beta_{res}^2 \theta_0^2}{2} \frac{k_\perp c p_{res}}{\omega_c} \Big(I_l (\Lambda) - I_l' (\Lambda) \Big) &
        \beta_{res}^2 I_l (\Lambda)
    \end{bmatrix}. \label{eq:epsA1}
\end{split}
\end{equation}
In a small-$\Lambda$ limit (taking a long perpendicular wave length limit with a finite $\theta_0$ but allowing an arbitrary wave length with a zero $\theta_0$ limit), $k_\perp^2 \ll \frac{2\omega_c^2}{c^2 p_{res}^2 \theta_0^2}$, its form is further simplified as
\begin{equation}
\begin{split}
    \overleftrightarrow{\epsilon}_A^{hot} = \pi \frac{\omega_p^2 n_{hot}}{\omega^2 n_{e}} \sum\limits_{l=-\infty}^\infty \frac{l^2 \omega_c^2 p_{res}\gamma^2_{res}}{k_\perp^2 c^2 (l \omega_c p_{res}- k_\| c)} &\Big( \omega \frac{dG}{dp}(p_{res}) - 2\frac{\omega-k_\|c \beta_{res}}{p_{res} \theta_0^2}G(p_{res}) \Big) \\
    &\times \exp(-\Lambda)I_l (\Lambda)
    \begin{bmatrix}
    1 & i \text{sgn}(l)^l & \frac{k_\perp c p_{res}}{l \omega_c} \\
    - i \text{sgn}(l)^l & 1 & - i \frac{k_\perp c p_{res}}{l \omega_c} \text{sgn}(l)^l \\
    \frac{k_\perp c p_{res}}{l \omega_c} & i \frac{k_\perp c p_{res}}{l \omega_c} \text{sgn}(l)^l & \frac{k_\perp^2 c^2 p_{res}^2}{l^2 \omega_c^2}
    \end{bmatrix}. \label{eq:epsA2}
\end{split}
\end{equation}
\end{widetext}
Equations \ref{eq:epsA1} and \ref{eq:epsA2} have a similar structure to the Maxwellian tensors (see Chapter 10 of Ref. \cite{Stix1992}).

\section{Waves emitted by runaway electrons}
\subsection{Avalanche-dominated momentum spectrum}
To evaluate ECE wave coefficients and kinetic instability drive, we specify $G(p) = \frac{\exp(-\frac{p}{p_0})}{p_0 p^2}$ as the exponential momentum spectrum with the average momentum $p_0$ that presumes a dominating runaway avalanche \cite{Rosenbluth1997NF, Aleynikov2015NF}. Then $G(p_{res})$ and $\frac{dG}{dp}(p_{res})$ are
\begin{equation}
    G(p_{res}) = \frac{\exp(-\frac{p_{res}}{p_0})}{p_0 p_{res}^2} \label{eq:E_spectrum}
\end{equation}
and
\begin{equation}
    \frac{dG}{dp}(p_{res}) = -\Big( \frac{1}{p_0} + \frac{2}{p_{res}} \Big)\frac{\exp(-\frac{p_{res}}{p_0})}{p_0 p_{res}^2}.
\end{equation}

This distribution would not reflect an exact energy spectrum in experiment due to neglects of other factors such as radiation and radial transport. The avalanche spectrum underestimates the low-energy, non-thermal population relative to the measured spectrum in DIII-D (see comparison shown by Fig. 13 in Ref. \cite{Hollmann2013NF}). Hence, we expect that it would yield a conservative estimate of the non-thermal ECE effect.

\subsection{Parameter selection for KSTAR-relevant conditions}
In this section, we apply the wave analysis to the full current-conversion scenario recently developed in KSTAR, similar to the TCV runaway plateau \cite{Decker2022NF}, which also exhibits $\mathrm{keV}$-scale non-thermal ECE signals. Because the pitch-angle spread and the momentum spectrum have not yet been measured in KSTAR, we first adopt the momentum spectrum given by Eq. \ref{eq:E_spectrum} and a conservative value of $\theta_0 = 0.15$, slightly smaller than the $\theta \approx 0.2$ inferred in DIII-D \cite{Hollmann2013NF}. After examining the resulting non-thermal ECE and the onset condition for kinetic instability, we then extend the analysis to higher $\theta_0\in(0.15, 0.5)$ within the Gaussian form. Incorporating the actual distribution will be important when the proposed model is validated against experiment, as done in Ref. \cite{Aleynikov2015NF}, but is not considered herein.

Detailed information on plasma parameters during the runaway plateau in KSTAR will be provided in a forthcoming publication. Here, we assume the standard major ($R_0 = 1.8 $ m) and minor radius ($a = 0.5$ m). The parameters are set to $B_T = 2.1$ T at $R=1.8$ m, $n_e = 1.5 \times 10^{19}$ m$^{-3}$, and $T_e = 3$ eV. The runaway electron density is estimated from a significant conversion of a plasma current of 350 kA, yielding $n_{RE} = 9 \times 10^{15}$ m$^{-3}$ whereas their characteristic momentum is assumed as $p_0=10$. We simplify a radial profile of $B_T$ as a 1/$R$-shape and other parameters as uniform.

\subsection{Non-thermal electron cyclotron emission\label{ssec:non-thermal}}
\subsubsection{Analytic spectral emissivity and absorption coefficient}
Let $j_\omega$ be the spectral emissivity quantifying the spontaneous emission rate and $\alpha_\omega$ be the absorption coefficient, given by the difference between the stimulated absorption rate and stimulated emission rate. For ECE interpretation emitted by disruption REs, we can split the wave coefficients $\alpha_\omega\approx\alpha_\omega^{th} + \alpha_\omega^{nth}$ and $j_\omega \approx j_\omega^{th} + j_\omega^{nth}$ by the thermal ($\alpha_\omega^{th}$, $j_\omega^{th}$) and non-thermal ($\alpha_\omega^{nth}$, $j_\omega^{nth}$) parts because the Maxwellian core with temperature in post-TQ phase is far from $p_0$ in phase space. For the thermal coefficients, we take the analytic ones given in Ref. \cite{Hutchinson2002,Rathgeber2013PPCF}.

From the cold plasma dispersion relation $[\vec{k} \vec{k} c^2 - \overleftrightarrow{I} k^2 c^2 + \omega^2 \overleftrightarrow{\varepsilon}_H] \cdot \vec{E}=0$, the solution corresponding to the X-mode wave yields the refractive index vector $\vec{N}\equiv c\vec{k}/\omega$ such that $N_\parallel \equiv \vec{N}\cdot \vec{B} / |\vec{B}|$ and $\vec{N}_\perp \equiv \vec{N} - N_\parallel \vec{B} / |\vec{B}|$,
\begin{equation}
    N_\perp = \sqrt{\frac{(\omega^2-\omega_p^2)^2 - \omega^2 \omega_c^2}{\omega^2(\omega^2 - \omega_p^2 - \omega_c^2)}}, \quad N_\parallel =0 
\end{equation}
and the polarization vector of electric field
\begin{equation}
    E_x = 1, E_y = - i \omega \frac{\omega^2 - \omega_p^2 - \omega_c^2}{\omega_c \omega_p^2}, E_z = 0.
\end{equation}

In tenuous media, the radiation transfer equation can be simply written as
\begin{equation}
    \frac{d}{ds} I_\omega = j_\omega - \alpha_\omega I_\omega \label{eq:rad_transf}
\end{equation}
where $I_\omega$ is the radiation intensity \cite{bekefi}.

By applying the Kirchhoff's radiation law and Poynting theorem \cite{bekefi}, the spectral emissivity and absorption coefficient can be found \cite{Lee2025phD}. The non-thermal spectral emissivity is
\begin{equation}
    j_\omega^{nth} = 2\pi \int dp_\| dp_\perp \Big[ p_\perp \eta_\omega f_{hot} \Big], \label{eq:j}
\end{equation}
where $\eta_\omega$ is the spectral particle emissivity
\begin{align}
    \eta_\omega \approx \sum\limits_{l=-\infty}^\infty \frac{m \omega^3 \omega_p^2}{8 \pi^2 c^2 k_\perp} \frac{\vec{E}^\dagger \cdot \overleftrightarrow{S}_l \cdot \vec{E}}{|E_y|^2} \delta(\omega - l \frac{\omega_c}{\gamma}).
    \label{eq:eta}
\end{align}
Using the results of Eqs. \ref{eq:Weber1}-\ref{eq:Weber6}, the resulting form becomes
\begin{equation}
    j_\omega^{nth} = \sum\limits_{l=-\infty}^\infty\frac{n_{hot}}{n_e} \frac{ml^2 \omega_c^2 \omega_p^2}{8\pi^2 c^2 k_\perp p_0 p_{res}}\exp(-\frac{p_{res}}{p_0}) \frac{\mathcal{J}_l}{|E_y|^2}, \label{eq:j1}
\end{equation}
where $\mathcal{J}_l$ is defined by
\begin{widetext}
\begin{equation}
\begin{split}
    \mathcal{J}_l \equiv \int &d\theta \theta (\vec{E}^\dagger \cdot\overleftrightarrow{S}_l \cdot \vec{E}) \frac{2}{\theta_0^2} \exp(- \frac{\theta^2}{\theta_0^2}) \\
    &=\exp(-\Lambda) \Big[I_l(\Lambda) \Big(\Big( \frac{l \omega_c}{\gamma_{res} k_\perp c}E_x - \frac{k_\perp c p_{res}\beta_{res}\theta_0^2}{2 \omega_c} i E_y )^2 -\Big(\Lambda + \frac{l^2}{\Lambda} \Big) \frac{\beta^2\theta_0^2}{2} E_y^2 \Big) \\
    &+ I_l'(\Lambda) \Big(il\beta_{res}^2 \theta_0^2 E_x E_y + \frac{k_\perp^2 c^2 p_{res}^2 \beta_{res}^2\theta_0^4}{2 \omega_c^2} E_y^2 \Big) \Big].
\end{split}
\end{equation}
\end{widetext}

The non-thermal absorption coefficient is
\begin{align}
    \alpha_\omega^{nth} &= -\frac{\omega^2}{c^2} \frac{\vec{E}^\dagger \cdot \overleftrightarrow{\varepsilon}_A^{hot} (f_0 = f_{hot}) \cdot \vec{E}}{k_\perp |E_y|^2} \nonumber \\
    &= - \frac{16\pi^4}{m\omega^3} \int dp_\| dp_\perp [\eta_\omega \gamma \omega U(f_0 = f_{hot})] \label{eq:alpha}
\end{align}
Substituting Eq. \ref{eq:epsA1} yields the analytic form of $\alpha_\omega^{nth}$
\begin{equation}
    \alpha_\omega^{nth} \approx \sum\limits_{l=-\infty}^\infty \frac{n_{hot}}{n_e} \frac{\pi l \omega_c \omega_p^2}{c^2 k_\perp \omega p_0 \beta_{res}^2}  \Big( \frac{1}{p_0} + \frac{2}{p_{res} \theta_0^2} \Big) \exp(-\frac{p_{res}}{p_0}) \frac{\mathcal{J}_l}{|E_y|^2} \label{eq:alpha1}
\end{equation}

For the ECE wave coefficients, the small-$\Lambda$ limit can be taken as shown in Eq. \ref{eq:epsA2}, which simplifies their forms,
\begin{align}
    j_\omega^{nth} &\approx \sum\limits_{l=-\infty}^\infty \frac{n_{hot}}{n_e} \frac{m l \omega^3 \omega_c \omega_p^2}{8 \pi^2 c^4 k_\perp^3 p_0 \beta_{res}} \nonumber \\
    &\times \exp\Big(-\Lambda - \frac{p_{res}}{p_0} \Big) I_l (\Lambda) \frac{(1+iE_y)^2}{|E_y|^2} \label{eq:j2}
\end{align}
and
\begin{align}
    \alpha_\omega^{nth} &\approx \sum\limits_{l=-\infty}^\infty \frac{n_{hot}}{n_e} \frac{\pi l\omega \omega_c \omega_p^2}{c^4 k_\perp^3 p_0 \beta_{res}^2} \Big( \frac{1}{p_0} + \frac{2}{p_{res}\theta_0^2} \Big) \nonumber \\
    &\times \exp\Big(-\Lambda - \frac{p_{res}}{p_0}\Big) I_l (\Lambda) \frac{(1+iE_y)^2}{|E_y|^2}. \label{eq:alpha2}
\end{align}

\subsubsection{Fictitious definition of local spectral non-thermal temperature in an infinite, homogeneous plasma\label{sssec:nth_homo_bb}}
Consider an infinite and homogeneous plasma in which magnetic field and non-thermal electron distribution are uniform in a configuration space and there is no Maxwellian population ($n_{hot}\to n_e$). This situation satisfies the radiative balance $dI_\omega / ds \to 0$ since the optical depth $\tau \equiv \int \alpha_\omega ds$ goes to an infinity. By analogue with black-body radiation in thermodynamic equilibrium, the spectral non-thermal temperature $T_{\omega}^{nth}$ can be fictitiously defined from the local source function $S_\omega \equiv j_\omega/\alpha_\omega$ as
\begin{equation}
    T_{\omega}^{nth} \, [\mathrm{eV}] = \frac{8\pi^2 c^2}{\omega^2e} \frac{j_\omega^{nth}}{\alpha_\omega^{nth}}  \approx \frac{mc^2}{e}\frac{p_{res}\beta_{res}\theta_0^2}{2}, \label{eq:T_nth}
\end{equation}
where we plugged Eqs. \ref{eq:j1} and \ref{eq:alpha1} (or \ref{eq:j2} and \ref{eq:alpha2}) into $j_\omega^{nth}$ and $\alpha_\omega^{nth}$, respectively, assumed that $l=2$ is the lowest mode number and neglected the higher contributions ($l\geq 3$). While the thermal temperature $T_e$ is constant in a $\omega$-space, the non-thermal temperature $T_{\omega}^{nth}$ is variant even under the spatial homogeneity. Instead, $T_{\omega}^{nth}$ is determined by the frequency ratio $\omega / \omega_c$ characterizing the resonant particle energy. This feature implies that when magnetic field is inhomogeneous $T_{\omega}^{nth}$ inferred from radiation localized at a single frequency $\omega$ reflects contributions accumulated across a range of particle energies. This in turn raises the question of how a non-thermal temperature should be fictitiously defined in a tokamak.

\begin{figure}
    \centering
    \includegraphics[width=\linewidth]{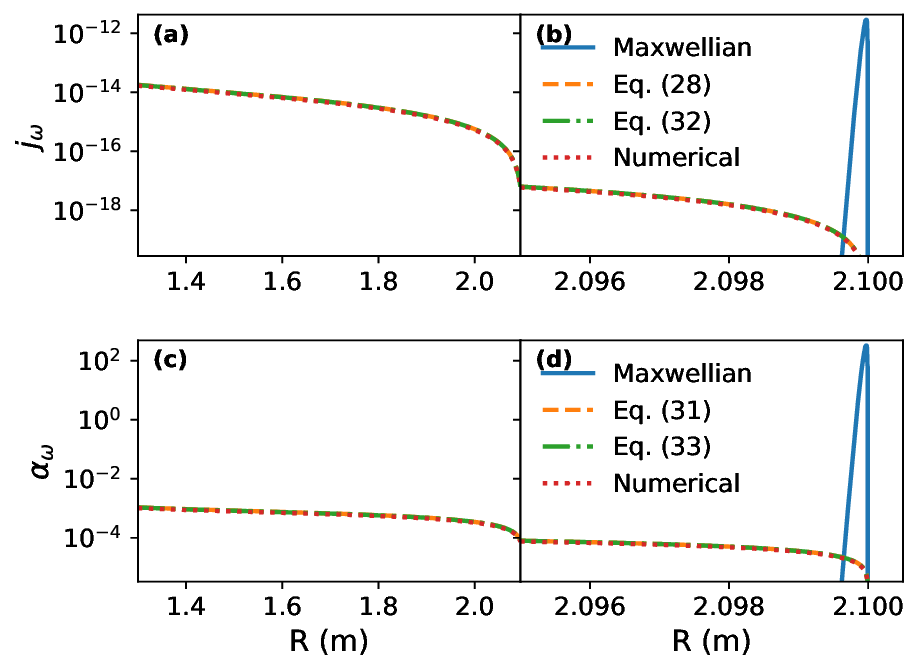}
    \caption{Radial profiles of $j_\omega$ (a,b) and $\alpha_\omega$ (c,d). Blue curve represents Maxwellian and other curves are hot electrons (REs). The generalized and small-$\Lambda$ limit solutions are orange-dashed and green-dashed-dotted, respectively. Red-dotted curve is obtained by numerical integration.}
    \label{fig:ECE_ja}
\end{figure}

\subsubsection{Numerical verification of $j_\omega^{nth}$ and $\alpha_\omega^{nth}$ under KSTAR-relevant condition}
To investigate non-thermal ECE in a tokamak, we developed the SYnthetic NOn-thermal electron cyclotron emission reconstruction tool (SYNO) code that computes the wave coefficients ($j_\omega$, $\alpha_\omega$) and solves the radiation transfer equation \ref{eq:rad_transf}. SYNO primarily uses the analytic forms of Eqs. \ref{eq:alpha1}, \ref{eq:alpha2}, \ref{eq:j1}, \ref{eq:j2}. But, it also includes a function that performs numerical integration of Eqs. \ref{eq:alpha}, \ref{eq:j}. Figure 1 shows the profiles of $j_\omega$ and $\alpha_\omega$ under conditions relevant to the KSTAR disruption runaway plateau, where $\omega$ corresponds to $2\omega_c$ at $R =2.1$ m. 

The analytic expressions (orange curve) are found to be in excellent agreement with the results obtained from numerical integration (red curve), as demonstrated in Fig. \ref{fig:ECE_ja}. Moreover, even the simplified form derived in the small-$\Lambda$ limit (green curve) retains a high level of accuracy. Indeed, we confirm that an error from the small angle approximations of $\sin \theta \approx \theta$ and $\cos \theta \approx 1$ are less than $10 \ \%$ for this specific example with $\theta_0 = 0.15$.

\begin{figure*}
    \centering
    \includegraphics[width=\linewidth]{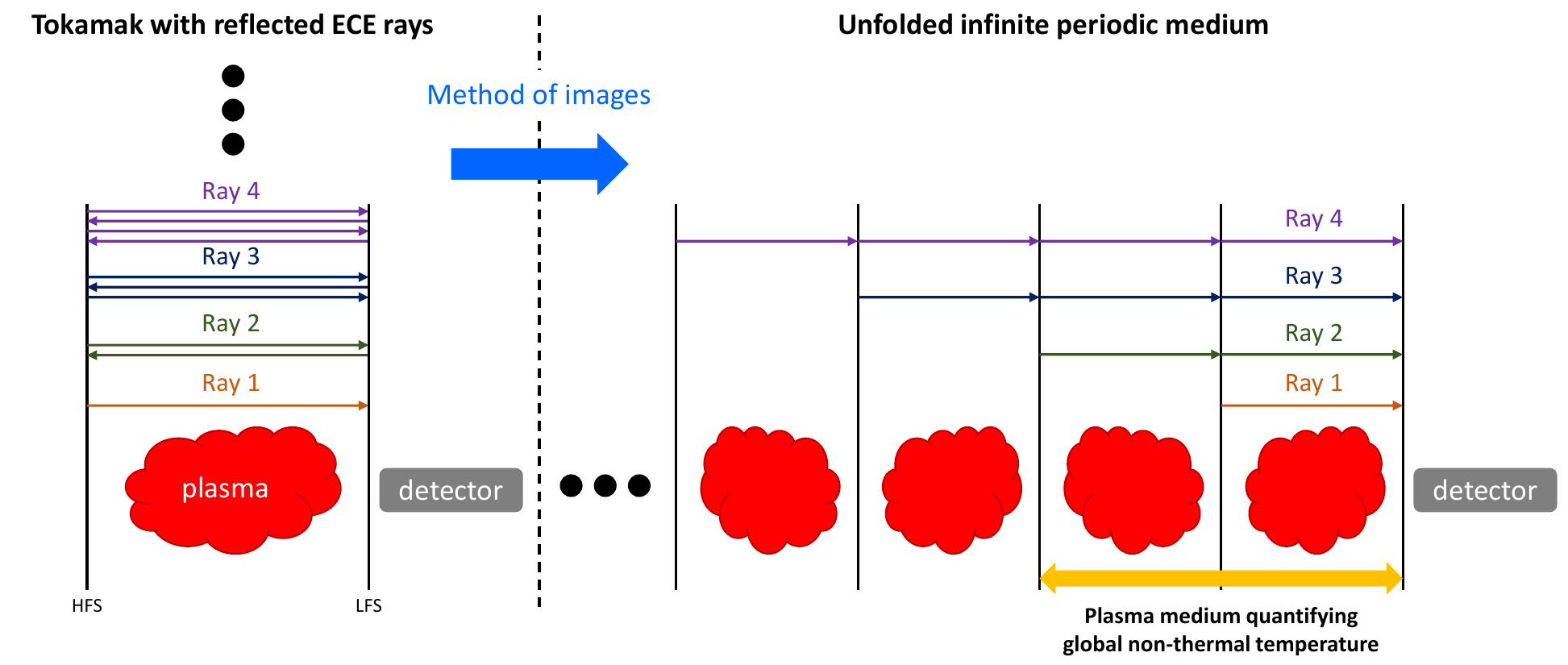}
    \caption{Schematic showing how the method of images unfolds the tokamak plasma medium.}
    \label{fig:schematic}
\end{figure*}

\subsubsection{Fictitious definition of global spectral non-thermal temperature in a tokamak}
As discussed in Sec. \ref{sssec:nth_homo_bb}, defining a representative non-thermal temperature in a tokamak is subtle owing to the spatial non-locality and the overlap of radiation from different energy ranges. We therefore introduce a fictitious definition of the spectral non-thermal temperature not as a "local" quantity, but as a "global" quantity from the viewpoint of an external observer looking at the plasma "medium" as a whole. Let the observer measure the temperature in a low field side (LFS) and then a trajectory of the measured ray from a high field side (HFS) traverses a local resonance layer of the core Maxwellian population before reaching the observer (a blue peak in Fig. \ref{fig:ECE_ja}). This region is spatially localized with a characteristic scale less than the order of $1$ mm but governs the ray absorption satisfying $\int \alpha_\omega^{th}ds \gg \int \alpha_\omega^{nth}ds$ due to a minority feature of hot populations $n_{hot}\ll n_{e}$. Solving Eq. \ref{eq:rad_transf} across this layer leads to
\begin{equation}
    I_\omega(\tau=\tau_{th}) \approx \frac{j_\omega^{th}}{\alpha_\omega^{th}} + \Big(I_\omega(\tau=0) -\frac{j_\omega^{th}}{\alpha_\omega^{th}}\Big) \exp(-\tau_{th}),
\end{equation}
where $\tau_{th}\equiv\int \alpha_\omega^{th}ds$ is the optical depth of the layer. In a $\tau_{th}\gg 1$ limit, however $I_\omega(0)$ is stronger $I_\omega$ goes to the black-body intensity corresponding to a local Maxwellian plasma with $T_e$. This suggests that the representative non-thermal temperature can be non-thermal only if the layer is optically thin enough. We refer to this layer as the absorption layer hereafter, which dominates the optical absorption and thereby determines if the target plasma medium can be non-thermal.

In an optically transparent tokamak, ECE rays pass through the absorption layer. The rays are reflected at the conducting wall boundary and the ECE detector measures a cumulative sum of all ray intensities. We apply the method of images to infinitely and periodically unfold the plasma medium as illustrated in Fig. \ref{fig:schematic}. We consider the complete reflection and neglect the X-O mode conversion for the formal definition of the global non-thermal ECE temperature in an idealized system; both of them are not a property of the plasma medium but features of the wall. In Sec. \ref{sssec:nth_homo_bb}, we showed that considering an infinite and homogeneous plasma makes the spectral non-thermal temperature $T_\omega^{nth}$ definable by taking a \textit{local} radiative-balance limit, i.e. $I_\omega\to S_\omega$, and then analogue with the black-body intensity. Similarly, the method of images suggests that the global spectral non-thermal temperature be also definable under the unfolded infinite periodic medium, as a \textit{global} radiative-balance limit of the optically transparent tokamak plasma medium. The global radiative balance condition is written by
\begin{equation}
    \int_{\text{plasma medium}} \frac{dI_\omega}{ds} ds = 0,
\end{equation}
where the integral boundary within the plasma medium is marked by a yellow double-headed arrow in Fig. \ref{fig:schematic}.

Let $I_\omega^+$ be the radiation intensity propagating from HFS to LFS obtained by solving Eq. \ref{eq:rad_transf} during the single pass and $I_\omega^-$ be that propagating from LFS to HFS. For $I_\omega^-$ to reach the detector, it needs at least one reflection at the HFS wall and experiences additional absorption during the travel across the tokamak. The resulting magnitude is $\exp(-\tau)I_\omega^-$, where $\tau\equiv \int \alpha ds$ measures the (single pass) optical depth between the conducting wall boundary. The reflection amplifies the radiation intensity by producing multiple rays that are accumulated at every successive double passes. The total radiation intensity becomes $(I_\omega^+ + \exp(-\tau)I_\omega^-) (1 + \exp(-2\tau) + \exp(-4\tau) + \cdots) = \frac{I_\omega^+ + \exp(-\tau)I_\omega^-}{1-\exp(-2\tau)}$. We fictitiously define the global spectral non-thermal temperature representing the tokamak plasma medium as
\begin{equation}
    T_{\omega}^{nth} \, [\mathrm{eV}] \equiv \frac{8\pi^2 c^2}{\omega^2e} \frac{I_\omega^+ + \exp(-\tau) I_\omega^-}{1-\exp(-2\tau)}. \label{eq:T_nth_global}
\end{equation}

With this definition, it becomes clear why strong ECE can arise during the runaway plateau phase without unstable kinetic instability. The reason is simply that, from the viewpoint of a horizontal ECE system, such a medium is not represented by a local thermal temperature but fictitiously characterized by a global non-thermal temperature with $T_{\omega}^{nth} \gg T_e$.

\subsubsection{Global spectral non-thermal temperature under KSTAR-relevant condition}
\begin{figure}
    \centering
    \includegraphics[width=\linewidth]{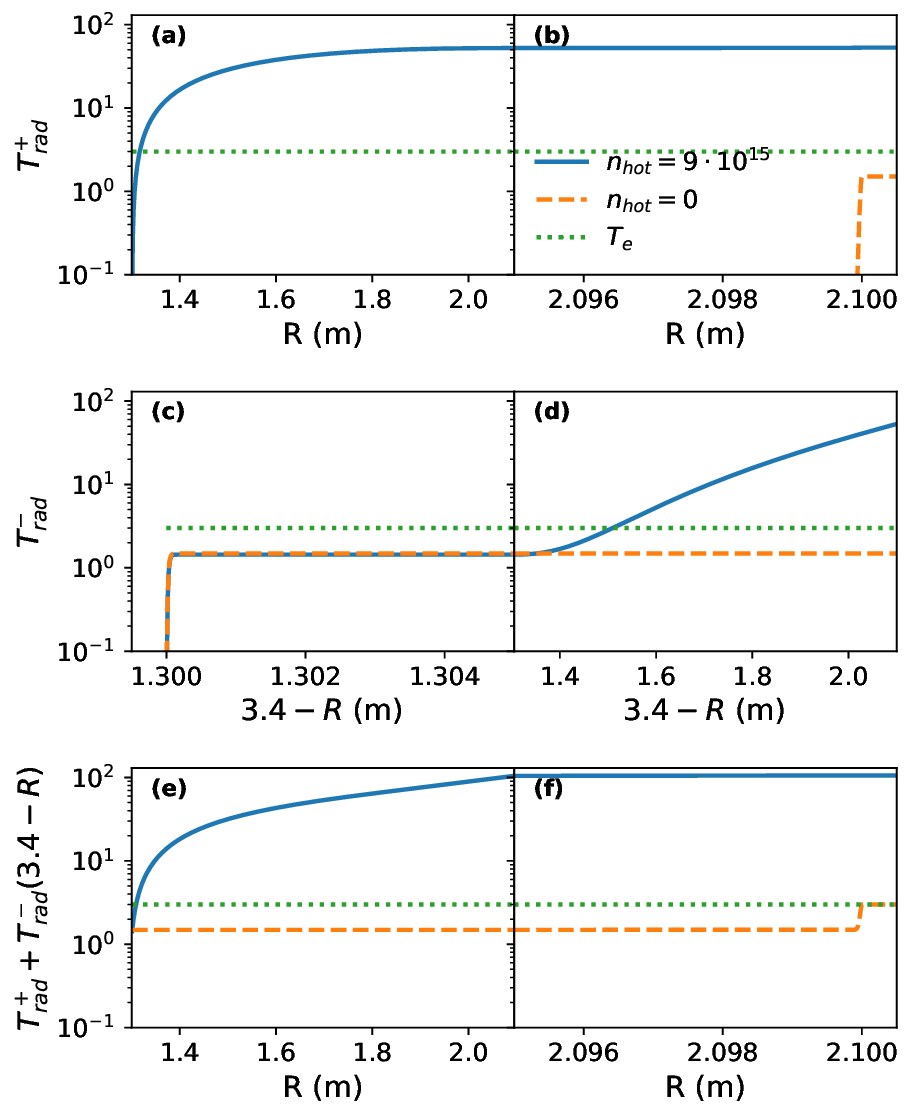}
    \caption{Ray-tracing simulation results performed by SYNO with $n_{hot}=9\times10^{15}m^{-3}$ (blue) and $n_{hot}=0$ (orange), respectively. Y-axis represents $T_{rad}^{+,-}$ in eV proportional to $I_\omega^{+,-}$ and their sum. Green curve shows the background electron temperature, i.e. $3$ eV.}
    \label{fig:Trad}
\end{figure}
In the runaway plateau phase, the optical layer can be thin enough due to a low companion plasma temperature. For a number of hot electrons, the resonance condition is satisfied across a spatially global region. Although $j_\omega$ associated with these electrons remains much lower than the peak value of the Maxwellian, their contribution outside the absorption layer can accumulate along the ray path toward the observer. This cumulative effect can produce a significant enhancement of the radiation intensity $I_\omega^+$ and $\exp(-\tau) I_\omega^-$, which can pass through the absorption layer and be multiplied after the wall reflections.

Under the same plasma condition with Fig. \ref{fig:ECE_ja}, we use SYNO to solve the radiation transfer equation \ref{eq:rad_transf} across the ray trajectories from $R=1.3$ m to $R=2.3$ m for $I_\omega^+$ and from $R=2.3$ m to $R=1.3$ m for $I_\omega^-$, respectively. The analytic wave coefficients given by Eqs. \ref{eq:j2} and \ref{eq:alpha2} are adopted in the small-$\Lambda$ limit. To separate contributions of rays with odd and even reflections, we accumulate their contribution in the radiative temperature form,
\begin{equation}
    T_{rad}^{+} \, [\mathrm{eV}] \equiv \frac{8\pi^2 c^2}{\omega^2e} \frac{I_\omega^+}{1-\exp(-2\tau)}
\end{equation}
and
\begin{equation}
    T_{rad}^{-} \, [\mathrm{eV}] \equiv \frac{8\pi^2 c^2}{\omega^2e} \frac{\exp(-\tau) I_\omega^-}{1-\exp(-2\tau)}.
\end{equation}

Figure \ref{fig:Trad} demonstrates the resulting $T_{rad}^{+}$ (a, b), $T_{rad}^{-}$ (c, d) and their sum (e, f). The sum of $T_{rad}^+(\mathrm{R}=2.1\, \mathrm{m}) + T_{rad}^-(\mathrm{R}=1.3\, \mathrm{m})$ yields $T_\omega^{nth}=106 \, \mathrm{eV}$. Since $I_\omega^-$ is calculated from LFS to HFS, we flip their propagation direction. Accordingly, we visualize Figs. \ref{fig:Trad}(c, d) with a reversed $x$-axis, and anti-symmetrically consider its argument in the sum in Figs. \ref{fig:Trad}(e, f). It is clarified that our definition of $T_\omega^{nth}$ in a tokamak (Eq. \ref{eq:T_nth_global}) well recovers the real electron temperature $T_e$ in a Maxwellian limit (orange curve). In the present of hot population, however, $T_\omega^{nth}$ reaches about $\mathcal{O}(100 \, \mathrm{eV})$ (blue curve) and far exceeds $T_e=3\,\mathrm{eV}$.

\begin{figure}
    \centering
    \includegraphics[width=\linewidth]{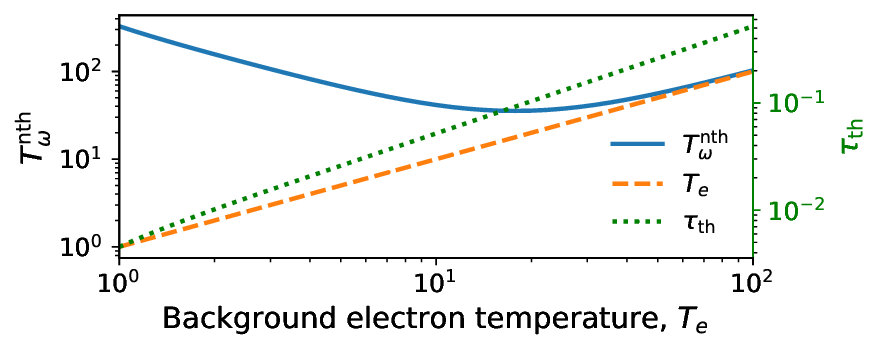}
    \caption{Evolution of $T_{\omega}^{nth}$ as a function of $T_e$ in eV. The right axis shows the optical depth $\tau_{\rm th}=\int \alpha_\omega^{th} ds$ across the absorption layer.}
    \label{fig:Te_scan}
\end{figure}

This finding is compatible with previous studies, in which a significant non-thermal effect in L-mode plasmas arises under strong quasi-linear diffusion at higher optical depth \cite{Harvey1993PoF, Liu2018NF}. Figure \ref{fig:Te_scan} demonstrates that the non-thermal effect originating from the small, highly anisotropic hot electrons gradually vanishes as the optical thickness across the absorption layer $\tau_{th}$ increases. Therefore, $T_\omega^{\rm nth}$ of an anisotropic medium can naturally exceed $T_e$ in the optically transparent runaway plateau, whereas quasi-linear diffusion or other additional mechanisms are likely required in a medium at flattop-level $T_e$.

\subsubsection{From global spectral non-thermal temperature to observed radiative temperature: incomplete wall reflection}
In a realistic tokamak, the wall reflection is incomplete and can involve the X-O mode conversion. This affects the radiative temperature $T_{rad}$ actually measured by a horizontal ECE system. Note that both of $T_\omega^{nth}$ and $T_{rad}$ are the fictitious temperatures. $T_{rad}$ is lower than $T_\omega^{nth}$ due to the incomplete reflective accumulation: $I_\omega$ does not reach the radiative balance. Let $r$ be the reflection coefficient and then $T_{rad}$ becomes
\begin{equation}
    T_{rad} \, [\mathrm{eV}] \equiv \frac{8\pi^2 c^2}{\omega^2e} \frac{I_\omega^+ + r\exp(-\tau) I_\omega^-}{1-r^2\exp(-2\tau)}. \label{eq:T_rad}
\end{equation}

The expression \ref{eq:T_rad} is somewhat different from Eq. A4 in Ref. \cite{Harvey1993PoF} because of considering the asymmetric contributions of $I_\omega^+$ and $I_\omega^-$. In inhomogeneous Maxwellian plasmas, $I_\omega^+ \approx I_\omega^-$ can be still a good approximation if the source function $S_\omega=j_\omega^{th}/\alpha_\omega^{th}$ can be locally invariant across the absorption layer. In the presence of sufficient hot populations, however, the source function varies globally. Within such a medium, the approximation $I_\omega^+ \approx I_\omega^-$ can be justified only if the optical depth is very thin: if it were, $I_\omega^+ \approx I_\omega^- \approx \int j_\omega ds$ would be met. It therefore appears that the optical thickness would have been small enough in Ref. \cite{Harvey1993PoF}, in which Eq. \ref{eq:T_rad} translates to $\lim_{\tau\to 0} T_{rad} \, [\mathrm{eV}] = \frac{8\pi^2 c^2}{\omega^2e} \frac{I_\omega^+}{1-r\exp(-\tau)}$.

Traditionally, practical rough estimates $r \geq 0.9$ had been obtained \cite{Bornatici1983NF}. Accordingly, in Ref. \cite{Harvey1993PoF}, they adopted $r=0.95$ for the FT-1 tokamak and $r=0.9$ for the DIII-D tokamak. In ASDEX-Upgrade, $r=0.99$ was set as a default \cite{Rathgeber2013PPCF}. Meanwhile, another dedicated study \cite{Austin1997} was conducted in DIII-D, which experimentally inferred a quite lower value of the reflection coefficient $r=0.74-0.77$ by employing optically-thin high harmonics of separate X and O waves and subsequently influenced a recent non-thermal ECE study \cite{Liu2018NF}. In KSTAR, there has been no convincing estimate of $r$. Hence, we set a plausible range of $r$ within $(0.7, 1)$, assuming that the wall reflectivity of KSTAR falls within the range reported for the devices discussed above. We do not treat the X-O mode conversion separately, presuming that its possible effect is effectively included in the conservative range of $r$.

\begin{figure}
    \centering
    \includegraphics[width=\linewidth]{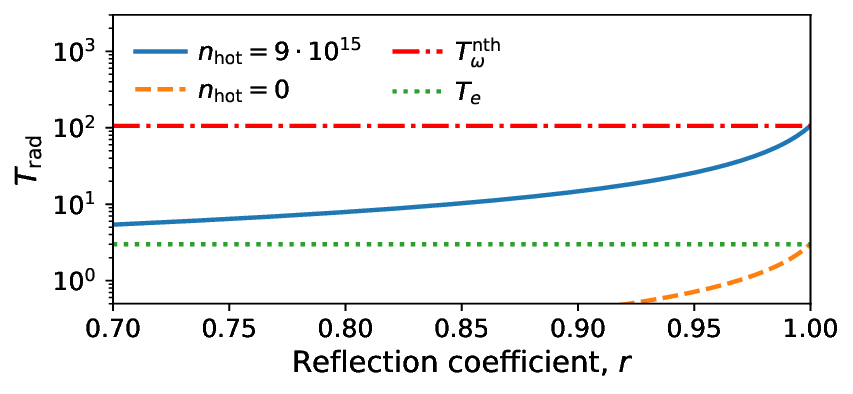}
    \caption{Evolution of $T_{rad}$ in eV as a function of $r$. Other settings are same to Figs. \ref{fig:ECE_ja} and \ref{fig:Trad}.}
    \label{fig:scan_reflection}
\end{figure}

Figure \ref{fig:scan_reflection} shows variation of $T_{rad}$ within the chosen range of $r \in (0.7, 1)$. This quantifies that although the reflection effect reduces $T_{rad}$, it is still higher than $T_e$. However, $T_{\rm rad}$ is found to drop rapidly as the reflection becomes less efficient; at $r=0.7$, it is reduced by an order of magnitude relative to $T_\omega^{\rm nth}$. This suggests that the actual $T_\omega^{\rm nth}$ in discharges where a few keV was observed could have been as high as few tens of keV.

\subsubsection{Beyond the analytic range of validity: numerical investigation of enhanced pitch-angle spread}
\begin{figure}
    \centering
    \includegraphics[width=\linewidth]{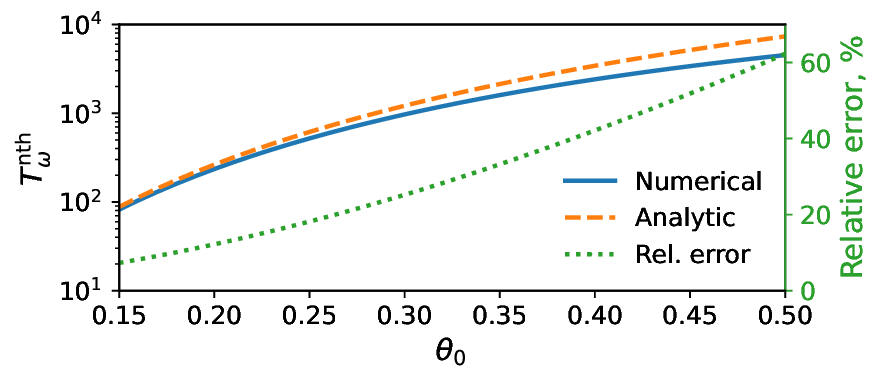}
    \caption{Evolution of $T_{\omega}^{nth}$ in eV as a function of $\theta_0$. Other settings are same to Figs. \ref{fig:ECE_ja} and \ref{fig:Trad}.}
    \label{fig:theta0_scan}
\end{figure}

\begin{figure*}
    \centering
    \includegraphics[width=\linewidth]{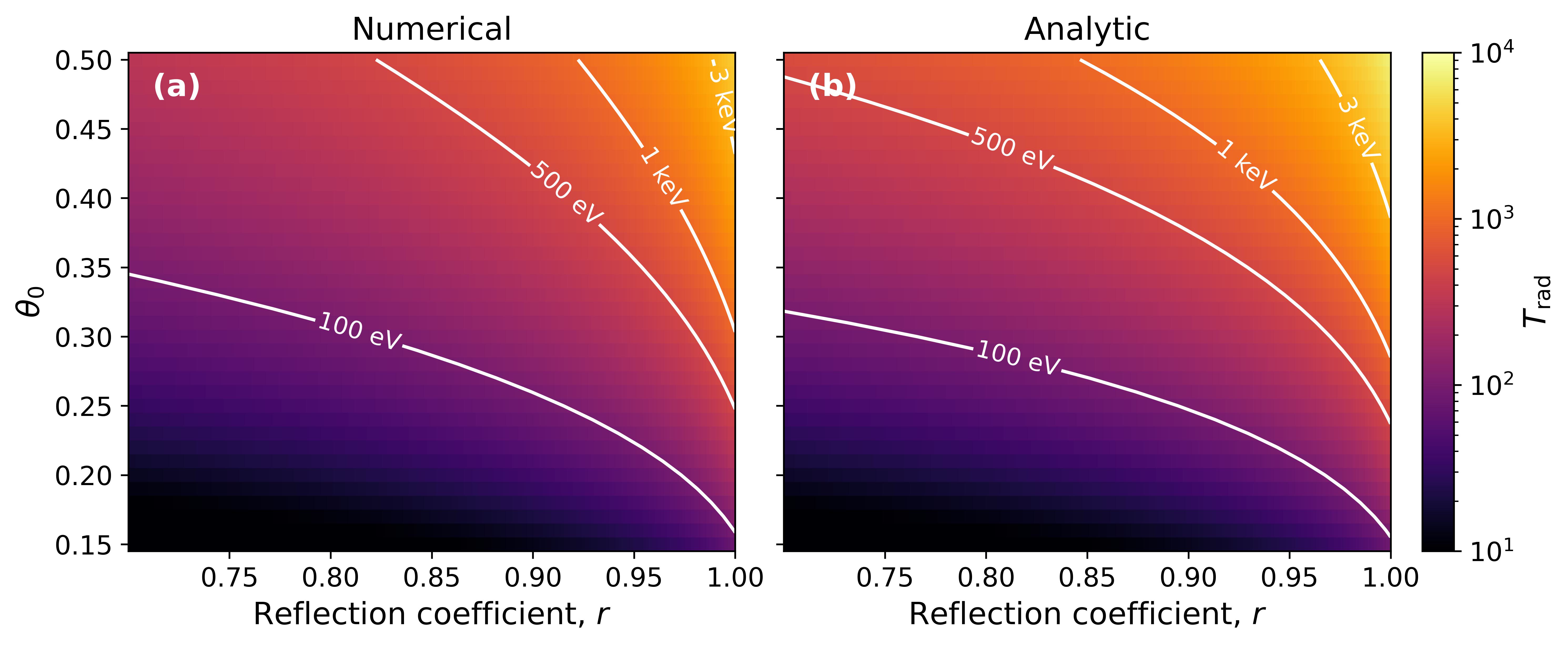}
    \caption{Contour plots of $T_{rad}$ in eV as a function of $\theta_0$ and $r$. Other settings are same to Figs. \ref{fig:ECE_ja} and \ref{fig:Trad}.}
    \label{fig:Trad_contour}
\end{figure*}

As will be analyzed in the next section \ref{ssec:kin_driv}, no kinetic instability is predicted at $\theta_0 = 0.15$ under the KSTAR-relevant condition. Accounting for the enhanced pitch-angle spread $\theta_0 \geq 0.15$ can amplify the non-thermal contribution of hot electrons, as discussed in Sec \ref{ssec:theta0_validity}. Figure \ref{fig:theta0_scan} demonstrates an effect of pitch-angle spread $\theta_0$ on $T_\omega^{nth}$. For the higher $\theta_0 \gtrsim 0.3$, the global spectral non-thermal temperature $T_\omega^{nth}$ can exceed $1\, \mathrm{keV}$. According to this result, observations of the keV-scale non-thermal ECE would be natural in a tokamak with highly reflective walls $r \approx 1$ once such a high $\theta_0$ was adequate to describe the pitch-angle spread of the radiating electrons. The adequacy, however, should be validated by independent kinetic simulations, which is clearly beyond the scope of this work.

Remind that our consideration of a realistic tokamak leaves huge uncertainty in $r\in (0.7, 1)$. Figure \ref{fig:Trad_contour}, however, supports the \textit{quantitative} statement that over $100 \, \mathrm{eV}$ non-thermal ECE would be explainable without unstable kinetic instability. The radiative temperature $T_{rad}$ of several hundred eV is predictable with the most conservative estimate of $r=0.7$ when $\theta_0 \geq 0.35$. However, explaining a keV-scale radiative temperature within the present model would require a high reflection efficiency $r \geq 0.9$. In conclusion, additional physics such as kinetic instabilities are not required to account for non-thermal ECE at the level of 100 eV to 1 keV across the parametric region between the corresponding contours.

Note that our analytic models of $j_\omega^{nth}$ and $\alpha_\omega^{nth}$ were developed with the small-$\theta$ approximation. Although these are still useful to comprehend a \textit{qualitative} trend, a great caution should be paid to make use of the numerical evaluations for quantifying the actual vales of $T_\omega^{nth}$ and $T_{rad}$ with $\theta_0 \gtrsim 0.2$ (see the relative error in Fig. \ref{fig:theta0_scan} and overestimation in $T_{rad}$ shown by the contour in Fig. \ref{fig:Trad_contour}(b)).

\subsection{Linear analysis of runaway-driven kinetic instability \label{ssec:kin_driv}}
\subsubsection{Analytic kinetic drive rate}
The cold plasma dispersion relation corresponding to the low frequency electron branch \cite{Akhiezer1975} yields the refractive index $N$
\begin{equation}
    N^2 = 1 - \frac{A}{B}. \label{eq:N+}
\end{equation}
where $A = 2\omega_p^2 (\omega^2 - \omega_p^2)$, $B=2\omega^2(\omega^2-\omega^2_p) - \omega^2\omega^2_c\sin^2\theta + \sqrt{\omega^4 \omega^4_c \sin^4\theta + 4\omega^2\omega^2_c (\omega^2-\omega^2_p)^2\cos^2\theta}$, $\cos \theta = N_\| / N$ and $\sin \theta = N_\perp / N$. The corresponding polarization vector of electric field is
\begin{equation}
    E_x = 1, E_y = i \frac{g_H}{\varepsilon_H - N^2}, E_z = - \frac{N_\| N_\perp}{\eta_H - N_\perp^2}.
\end{equation}

Linear kinetic instability growth \cite{Aleynikov2015NF} is written as
\begin{equation}
    \gamma^{kin}_{net} = \gamma^{kin}_{drive} - \gamma_{damp}^{coll}
\end{equation}
where $\gamma^{kin}_{net}$ is the net growth rate, $\gamma^{kin}_{drive}$ is the kinetic drive given by
\begin{equation}
    \gamma_{drive}^{kin} \equiv - \frac{\omega^2\vec{E}^\dagger \cdot \overleftrightarrow{\varepsilon}_A^{hot} \cdot \vec{E}}{\vec{E}^\dagger \cdot \frac{\partial}{\partial \omega}(\omega^2 \overleftrightarrow{\varepsilon}_H) \cdot \vec{E}} \label{eq:gamma_kin}
\end{equation}
and $\gamma_{damp}^{coll}$ is the collisional damping rate given by
\begin{equation}
    \gamma_{damp}^{coll} \equiv \frac{\omega^2\vec{E}^\dagger \cdot \overleftrightarrow{\varepsilon}_A^{coll} \cdot \vec{E}}{\vec{E}^\dagger \cdot \frac{\partial}{\partial \omega}(\omega^2 \overleftrightarrow{\varepsilon}_H) \cdot \vec{E}}. 
\end{equation}

\begin{widetext}
The analytic solution of kinetic drive is 
\begin{equation}
    \gamma_{drive}^{kin} = \frac{1}{\vec{E}^\dagger \cdot \frac{\partial}{\partial \omega}(\omega^2 \overleftrightarrow{\varepsilon}_H) \cdot \vec{E}} \sum\limits_{l=-\infty}^\infty \frac{\pi \omega_p^2 n_{hot} \gamma^4_{res}}{n_e(l \omega_c p_{res} - k_\| c)p_0 p_{res}} \Big( \omega (\frac{1}{p_0} + \frac{2}{p_{res}}) + 2\frac{\omega-k_\|c \beta_{res}}{p_{res} \theta_0^2} \Big) \exp(-\frac{p_{res}}{p_0}) \mathcal{S}_l  \label{eq:gamma_kin1}
\end{equation}
where $\mathcal{S}_l$ is defined by
\begin{equation}
\begin{split}
    \mathcal{S}_l \equiv \int &d\theta \theta (\vec{E}^\dagger \cdot\overleftrightarrow{S}_l \cdot \vec{E}) \frac{2}{\theta_0^2} \exp(- \frac{\theta^2}{\theta_0^2}) \\
    &=\exp(-\Lambda) \Big[I_l(\Lambda) \Big(\Big( \frac{l \omega_c}{\gamma_{res} k_\perp c}E_x - \frac{k_\perp c p_{res}\beta_{res}\theta_0^2}{2 \omega_c} i E_y + \beta_{res} E_z \Big)^2 - \Big(\Lambda + \frac{l^2}{\Lambda} \Big) \frac{\beta^2\theta_0^2}{2} E_y^2 \Big) \\
    &+ I_l'(\Lambda) \Big(il\beta_{res}^2 \theta_0^2 E_x E_y + \frac{k_\perp^2 c^2 p_{res}^2 \beta_{res}^2\theta_0^4}{2 \omega_c^2} E_y^2  + i\frac{k_\perp c p_{res}}{\omega_c} \beta_{res}^2 \theta_0^2 E_y E_z \Big) \Big].
\end{split}
\end{equation}
In the small-$\Lambda$ limit, it reduces to
\begin{equation}
\begin{split}
     \gamma_{drive}^{kin} &= \frac{1}{\vec{E}^\dagger \cdot \frac{\partial}{\partial \omega}(\omega^2 \overleftrightarrow{\varepsilon}_H) \cdot \vec{E}} \sum\limits_{l=-\infty}^\infty \frac{\pi n_{hot} \omega_p^2 \omega_c^2 \gamma_{res}^2}{n_e (l \omega_c p_{res} - k_\| c) k_\perp^2 c^2 p_0 p_{res} }\Big( (\frac{1}{p_0} + \frac{2}{p_{res}})\omega + \frac{\omega - k_\| c \beta_{res}}{p_{res}} \frac{2}{\theta_0^2}\Big) \\
    &\times \Big(l(1+i(-1)^l E_y) + \frac{k_\perp c p_{res}}{\omega_c}E_z \Big)^2 \exp\Big(-\frac{p_{res}}{p_0}-\Lambda \Big) I_l (\Lambda) \label{eq:gamma_kin2}
\end{split}
\end{equation}
\end{widetext}
In Ref. \cite{Aleynikov2015NF}, the analytic kinetic drive expression was presented under the anomalous Doppler resonance condition. Our formula recovers this by taking the small-$\theta_0$ limit and considering only the $l=-1$ resonance, which yields
\begin{equation}
     \gamma_{drive}^{kin} = \frac{\pi}{2e} \frac{n_{hot} \omega_p^2}{n_e p_{0}} \frac{(-1 + iE_y + \frac{k_\perp c p_{res}}{\omega_c}E_z)^2}{\vec{E}^\dagger \cdot \frac{\partial}{\partial \omega}(\omega^2 \overleftrightarrow{\varepsilon}_H) \cdot \vec{E}} \frac{\omega_c \gamma_{res}}{k_\| c + \omega_c p_{res}}.
\end{equation}
This equation is identical to Eq. (30) of Ref. \cite{Aleynikov2015NF} once we take $\omega_c \to -\omega_c$ to match the sign convention and use $p_{0}\approx \gamma_{res}$ corresponding to Eq. (29) of Ref. \cite{Aleynikov2015NF}.

\subsubsection{Numerical verification of $\gamma_{drive}^{kin}$ under KSTAR-relevant condition}
\begin{figure}
    \centering
    \includegraphics[width=\linewidth]{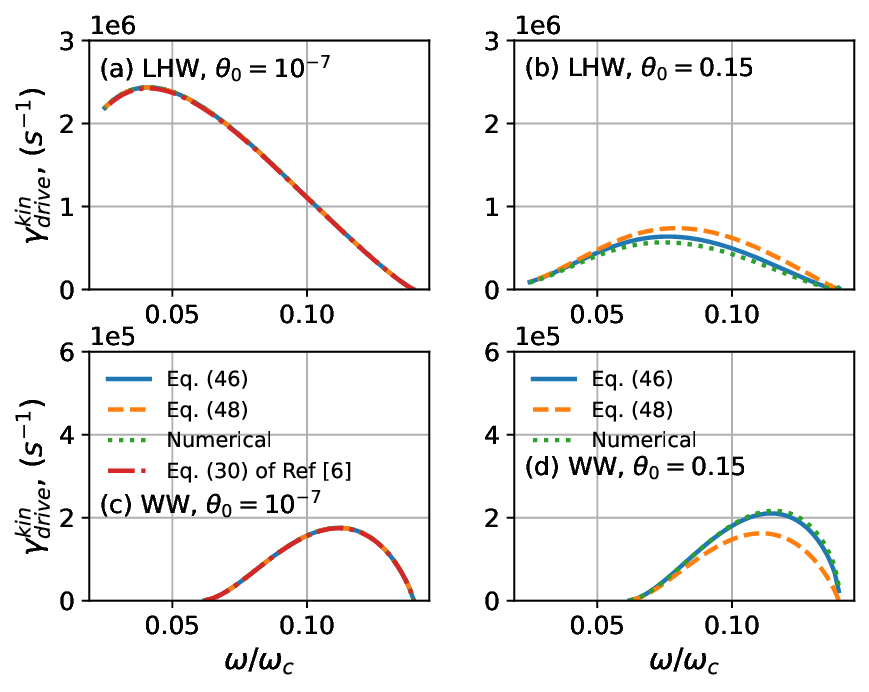}
    \caption{$\gamma_{drive}^{kin}$ for LHW (a,b) and WW (c,d) branches with $\theta_0=10^{-7}$ (a,c) and $\theta_0=0.15$, respectively. Plasma parameters are same to Fig. \ref{fig:ECE_ja} at $R=1.8$ m. Blue-solid and orange-dashed curves are analytically obtained using Eqs. \ref{eq:gamma_kin1} and \ref{eq:gamma_kin2}, respectively. Green-dotted curve is a numerical result of Eq. \ref{eq:gamma_kin} using KIAT. Red-dashed-dotted curve is analytic formula in a $\theta_0 \to 0$ limit, presented in Ref. \cite{Aleynikov2015NF}.}
    \label{fig:kin_drive}
\end{figure}
We developed the Kinetic Instability Analysis Tool (KIAT) code. It routinely detects wave branches resonant with REs of momentum $p_0$ and zero Larmor radius and computes linear growth rate. Similar to SYNO, KIAT includes a function that performs numerical integration of Eq. \ref{eq:gamma_kin}. Figure \ref{fig:kin_drive} shows the analytic $\gamma_{drive}^{kin}$ (\ref{eq:gamma_kin1}) (blue curve) agrees well with the results from numerical integration (green curve), where we only consider the lower hybrid wave (LHW or magnetized plasma wave) and whistler wave (WW) and anomalous Doppler resonance ($l=-1$). The maximum absolute errors are $7.4\times10^{4} s^{-1}$ and $1.5\times10^{4} s^{-1}$ for LHW and WW, respectively. However, we found that the proper $\theta_0$ to adopt the small-$\Lambda$ limit (\ref{eq:gamma_kin2}) is narrower than the non-thermal ECE addressed in Sec. \ref{ssec:non-thermal}. Note that Eq. \ref{eq:gamma_kin2} is valid in a small-$\theta_0$ limit, reproducing Eq. (30) of Ref. \cite{Aleynikov2015NF}. Our analytic formula (\ref{eq:gamma_kin1}) produces $\gamma_{drive}^{kin}$ slightly higher than the values in Fig. 3 of Ref. \cite{Aleynikov2015NF} as reported in Ref. \cite{Zhang2025PoP}. It appears to have a typo during the numerical implementation of the correct formulation (Eq. (21) in Ref. \cite{Aleynikov2015NF}). This hopefully answers to a question raised in Ref. \cite{Zhang2025PoP}, "It remains unclear what may have contributed to the discrepancy between our results, which were obtained separately from analytical analysis and numerical integration, and those computed in Ref. 19", where Ref. 19 corresponds to Ref. \cite{Aleynikov2015NF} in our work.

\subsubsection{Linear stability analysis under KSTAR-relevant condition}
\begin{figure}
    \centering
    \includegraphics[width=\linewidth]{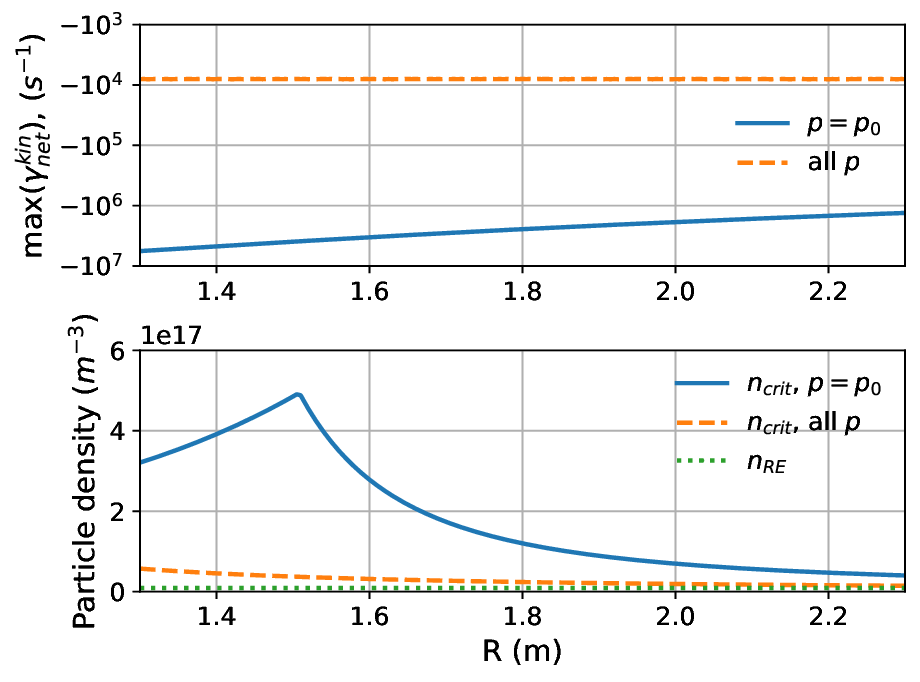}
    \caption{Radial profiles of $\max(\gamma_{net}^{kin})$ in top panel, $n_{crit}$ (blue solid and orange dashed) and $n_{RE}$ (green dotted) in bottom panel. In blue, the wave branches resonating with $p=p_0$ and zero Larmor radius are considered whereas in orange all waves are considered. Plasma parameters are same to Fig. \ref{fig:ECE_ja}.}
    \label{fig:gamma_max}
\end{figure}
Let $n_{crit}$ be the critical density of kinetic instability onset such that $\gamma_{net}^{kin}(n_{hot}=n_{RE}) = 0$. We perform a linear stability analysis on the WW and LHW branches shown by Fig. \ref{fig:kin_drive} (b,d) due to anomalous Doppler resonance by using KIAT with the analytic kinetic drive \ref{eq:gamma_kin1} to compute $\max(\gamma_{net}^{kin})$ and $n_{crit}$. Figure \ref{fig:gamma_max} shows that the instability onset is forbidden in the parameters used in Figs. \ref{fig:ECE_ja} and \ref{fig:Trad}, since $n_{RE}$ (green curve) does not reach $n_{crit}$ (blue curve) and the net growth rate is negative across all radial points. This suggests that if the assumption of $f_{hot}$ (Eq. \ref{eq:f}) and the estimate of $\theta_0$ are reasonable, such a distribution wouldn’t collapse under unstable waves, but could instead remain stable.

For highly energetic particles, additional stopping power arises from synchrotron and bremsstrahlung radiation, together with a significant orbit shift that limits the maximum energy \cite{Knoepfel1979NF}. The energy spectrum described by Eq. \ref{eq:E_spectrum} may therefore overestimate the number of ultra-relativistic electrons. Nevertheless, we confirm that the maximum growth rate over all waves including those resonant with such energetic electrons remains negative as clarified by the orange curve.

\subsubsection{A possible but neglected candidate}
Although the parametric decay of the slow-X mode may occur \cite{Zhang2026PRE}, we neglect it in the present analysis because its impact on ECE observations remains unclear, rather than due to any lack of physical relevance.

\section{Discussion}
In L-mode plasmas, previous studies about non-thermal ECE signals measured by a horizontal ECE system have often attribute the radiative temperature increment to wave-particle interactions driven by kinetic instabilities or by externally injected waves \cite{Harvey1993PoF, Liu2018NF}. Observations of strong non-thermal ECE during the runaway plateau phase, however, suggest that temperature anomalies be explainable without them \cite{Hollmann2013NF, Aleynikov2015NF}. In an optically transparent tokamak, a plasma medium consists of a major isotropic cold population and a minor anisotropic hot population. We propose that characterizing such a medium necessarily calls for the fictitious definition of the global spectral non-thermal temperature in analogue with an infinite, homogeneous medium. Under this definition, a tokamak plasma medium during the runaway plateau can be sufficiently non-thermal. Indeed, our simplified analysis shows that once the pitch-angle spread exceeds a modest level ($\theta_0 \geq 0.35$), a radiative temperature of several hundred eV can arise for KSTAR-relevant plasma parameters without invoking kinetic instabilities or injecting external waves.

This finding is compatible with the earlier studies \cite{Harvey1993PoF, Liu2018NF} rather than contradicting them. The global non-thermal temperature depends strongly on the optical thickness of the absorption layer, and we find that the non-thermal feature disappears once this thickness exceeds a certain level. The two pictures are thus distinguished by parametric regimes. In L-mode plasmas, the Maxwellian background renders the medium moderately optically thin, so that a non-thermal feature requires a more pronounced variation of the distribution beyond a standard anisotropic form. In the runaway plateau phase, by contrast, the low background temperature drives the plasma medium toward a optically transparent limit, where even the small contribution of the anisotropic medium can result in temperature anomalies.

Although the Gaussian pitch-angle distribution adopted in this study may differ from the actual runaway electron distribution, it is sufficient for qualitatively exploring an origin of the emergence of non-thermal ECE in the post-disruption phase. A quantitative validation, however, necessitates a realistic particle distribution from measurements or kinetic simulations, and a dedicated validation against the measured ECE spectrum, which are future works of this work. In this work, we presented an analytical hot plasma dielectric tensor and obtained compact analytic expressions for the corresponding non-thermal wave coefficients and the linear kinetic drive. To the best of our knowledge, the former is presented here for the first time in the context of non-thermal ECE modeling, while the latter represents a finite-$\theta_0$ correction of Eq. (30) found in Ref. \cite{Aleynikov2015NF}. Within the analytical framework developed here, applying the non-thermal ECE model in regimes where the onset of linear kinetic instability is not predicted is self-consistent, given that both descriptions originate from the same hot plasma dielectric tensor.

\begin{acknowledgments}
Y. Lee acknowledges Dr. Pavel Aleynikov and Dr. Qile Zhang for assistance with the verification of KIAT and for helpful comments on the interpretation of the finite-$\theta_0$ correction. The authors are grateful to Dr. Jayhyun Kim, Dr. Jaemin Kwon, Ms. Hyelin Kang, Dr. Dong-Kwon Kim and Dr. Kyu-Dong Lee for useful suggestions. This research was supported by R\&D Program of "Optimal Basic Design of DEMO Fusion Reactor, CN2602-4" through the Korea Institute of Fusion Energy (KFE) funded by the Government funds.
\end{acknowledgments}

\section*{Author Contributions}
\textbf{Yeongsun Lee}: Conceptualization (lead); Formal analysis (lead); Investigation (equal); Methodology (lead); Validation (lead); Writing - original draft (lead); Writing - review \& editing (equal). 
\textbf{Kikyung Park}: Formal analysis (equal); Investigation (support); Methodology (support); Validation (equal); Writing - review \& editing (equal). 
\textbf{Tchanou Park}: Formal analysis (equal); Investigation (support); Methodology (support); Validation (equal); Writing - review \& editing (equal). 
\textbf{Gunsu Yun}: Investigation (support); Methodology (support); Supervision (support); Writing - review \& editing (equal). 
\textbf{Yong-Su Na}: Funding acquisition (supporting); Supervision (equal); Writing - review \& editing (equal). 
\textbf{Jong-Kyu Park}: Conceptualization (equal); Funding acquisition (lead); Project administration (lead); Supervision (lead); Writing - review \& editing (equal).

\appendix

\bibliographystyle{unsrt}
\bibliography{references}

\end{document}